# A Deep Learning-Augmented Stand-off Radar Scheme for Rapidly Detecting Tree Defects

Jiwei Qian, *Student Member, IEEE*, Yee Hui Lee, *Senior Member, IEEE*, Kaixuan Cheng, *Student Member, IEEE*, Qiqi Dai, *Student Member, IEEE*, Mohamed Lokman Mohd Yusof, Daryl Lee, and Abdulkadir C. Yucel, *Senior Member, IEEE*

*Abstract*—Tree defect detection is crucial for the structural health screening of trees. Existing nondestructive testing (NDT) techniques for tree defect detection require time-consuming and labor-intensive measurement campaigns. This discourages their application for the routine structural health screening of whole populations of managed urban trees. To address this issue, this study proposes a deep-learning augmented stand-off radar scheme for contactless scanning of tree trunks and rapid detection of tree defects. In this scheme, the antenna is moved along a straight trajectory at a distance from the tree trunk to obtain the trunk's B-scan. The obtained raw B-scan is then processed by a signal-processing framework specifically developed for revealing the scattering signatures of defects in B-scan, which achieves a 30 dB and 22 dB increase in the signal-to-clutter and noise ratio of the measurement data of tree trunk samples and living trees, respectively. Finally, the processed B-scan is input into a multilevel feature fusion neural network particularly designed for extracting the signature of the defect in the processed B-scan in real time. The developed scheme's applications to the detection of defects in real fresh-cut tree trunks show that the stand-off radar scheme can detect tree defects with 96% accuracy. This stand-off radar scheme is the first contactless NDT technique for tree defect detection while operated on a straight trajectory and potentially can be integrated into the routine tree inspection workflow which is part of urban tree management.

## I. INTRODUCTION

URBAN trees are "living structures" that provide significant ecosystem services and benefits like shade, improved air quality, noise control, and beautification of the environment, etc. These benefits increase as the age and size of the trees increase. However, as trees get older and larger, they are more likely to develop defects in the form of wood decay or others that potentially increase the risk of whole tree or tree part failures which can pose a danger to people and property. Regular and routine structural health screening of trees is important to reduce the risk of these failures. Ideally, such regular structural health screenings should provide accurate and timely detection of the internal defect (e.g., cavity and decay) that is often hidden from the routine visual inspection process [1]. The screening may potentially provide the required information like, defect type, severity, and how far the defect has progressed along the length of the tree part, this information can be used to assess the degradation in the strength properties of the investigated tree/tree part to warrant management intervention [2].

Current nondestructive testing (NDT) or minimally invasive techniques for tree defect detection include sonic tomography [3] and electrical resistivity tomography [4]. These techniques require a large number of sensors, are labor-intensive, and necessitate significant time to perform measurements on each tree. Besides that, microwave tomography has been proposed to detect tree defects by imaging the distribution of dielectric properties of the tree under test [5], [6]. In particular, the electromagnetic fields scattered along all directions from the target tree are obtained by either a circularly moved receiver [7] or an array of receiving antennas [8]. Then, the nonlinear and ill-posed inverse scattering problem is solved by iterative methods [8], [9]. In general, such methods are more computationally intensive than radar-based approaches, and therefore more challenging to be applied to the real-time or rapid detection of defects required for routine structural assessments of trees.

Recently, ground-penetrating radar (GPR), which comprises a single compact sensing module, has become an attractive candidate for detecting tree defects because of its simplicity [10], [11] and low computational time requirement for imaging the tree defects [12], [13]. However, when it comes to defect detection for the entire urban tree population, the GPR technique becomes impractical in two aspects. First, GPR requires the contact of the antenna with the tree trunk surface while moving the antenna along the perimeter plane slowly to collect consistent data [14], [9]. Such a data collection process is time-consuming [15], which rules out its application to the health screening of whole tree populations. Although augmented positioning systems, such as the binary ruler [16] and the wheel-measuring device [12], can be used to accelerate the process, the antenna movement is still to be performed slowly to avoid measurement error induced by surface roughness of the trunk. Additionally, current GPR techniques

This work was supported by National Parks Board, Singapore. *(Corresponding author: Yee Hui Lee, Abdulkadir C. Yucel.)*

Jiwei Qian, Yee Hui Lee, Kaixuan Cheng, Qiqi Dai, and Abdulkadir C. Yucel are with the School of Electrical and Electronic Engineering, Nanyang Technological University, Singapore 639798 (e-mail: qian0069@e.ntu.edu.sg; eyhlee@ntu.edu.sg; chen1519@e.ntu.edu.sg; daiq0004@e.ntu.edu.sg; acyucel@ntu.edu.sg).

Mohamed Lokman Mohd Yusof, and Daryl Lee are with the Centre for Urban Greenery and Ecology, National Parks Board, Singapore 259569 (e-mail: mohamed_lokman_mohd_yusof@nparks.gov.sg, daryl_lee@nparks.gov.sg).

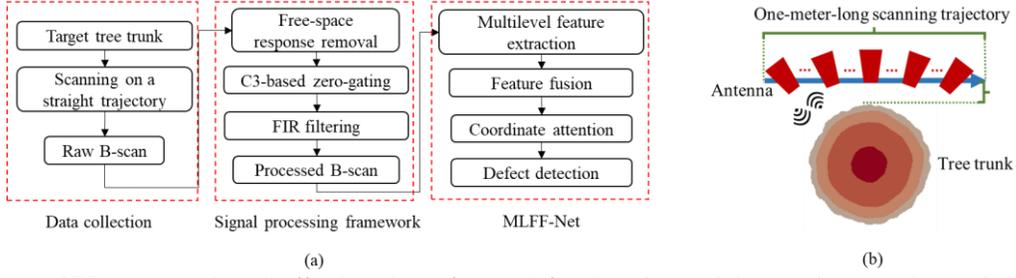

Fig. 1. (a) Proposed DL-augmented stand-off radar scheme for tree defect detection, and (b) scanning a tree by moving the antenna on a straight trajectory.

rely upon imaging techniques, including modified Kirchhoff migration [12] and matching pursuit algorithm [17], used to identify the defect region inside the tree trunk [18]. The accuracy of these techniques depends on "clean and continuous signatures" of defects in the measurement, which are not always retained after necessary signal processing steps [12]. Full-waveform inversion [19] can image the permittivity distribution of the investigated region accurately. However, it is computationally costly as it requires the repeated and iterative execution of the forward solver and hence lacks of providing quick assessments of trees. To overcome these shortcomings in current GPR technology, a novel stand-off radar scheme for rapidly detecting tree defects is needed for practical routine field application.

In areas other than tree defect detection such as road inspection [20], landmine detection [21], and cultural heritage [22], GPR systems can be operated in the stand-off configuration and used to detect the subsurface objects with the antenna being a few wavelengths away from the medium under investigation. By mounting the antennas on the wheel-based platforms along a straight trajectory, the time for screening target media can be reduced to minutes [23]. To the best of our knowledge, such a stand-off straight scanning trajectory has never been used to detect trees' internal defects. In the past, a stand-off circular scanning prototype is developed to measure multi-view scattering information of tree trunk samples with high efficiency [7] [24]. In this approach, by fixing the transmitting antenna and moving the receiving antenna in a circular trajectory around the tree trunk sample, the multi-view scattering information is collected in a contactless way, which is further utilized to image the tree defect by the computationally costly microwave tomography algorithms [6]. However, such a circular scanning configuration requires more time to detect defects in living trees compared to a straight scanning trajectory. The realization of a stand-off radar system that can work in a contactless straight scanning configuration requires the design of a new antenna with ultrawide bandwidth, high gain, compact size, and narrow beamwidth to maximize the power penetration into the tree trunk [25]. Apart from this hardware requirement, there also exists software requirements for processing and interpreting the signals reflected from tree defects, as below.

1. New signal processing techniques are needed to remove the undesired clutter due to the reflections from the air-bark

interface so that signals reflected from defects can be distinguished in the two-dimensional radar chart, which is known as the B-scan. Since the straight trajectory is not conformal to the side surface of the tree, the signal reflected from the bark appears as a hyperbolic curve in the B-scan, which cannot be eliminated by existing clutter removal techniques, including background removal [9] and singular value decomposition (SVD) [11].

2. A new deep learning (DL) technique for distinguishing defect signatures in noisy measurement data is required. While DL techniques have been used for hyperbola recognition [27], object classification [28], parameter estimation [29] [30], and subsurface imaging [31] [32] for a wide range of GPR applications, most of the training samples are either measurement data of the objects buried in a homogeneous medium [31] or B-scans with clear signatures of targets [33], which are not applicable for tree defect detection. Moreover, limited studies deploy DL methods to detect tree defects. Although [34] considers the layered structure of trees for imaging and classification of tree defects, the network is only tested with clean synthetic data and cannot be accurate for noisy measurement data performed in the field.

In this study, we propose a DL-augmented stand-off radar scheme for rapid defect detection during the health monitoring of trees. The proposed scheme, described in Fig. 1(a), consists of three parts. First, measurements are conducted on a one-meter-long straight trajectory at a distance from the bark [Fig. 1(b)]. A developed stepped-frequency continuous-wave (SFCW) stand-off radar system with a novel antenna [25] records one reflected signal (A-scan) from the tree trunk with each two-centimeter translation of the antenna and then forms the B-scan of the tree trunk by spatially stacking A-scans together. With a specially designed automatic motorized antenna slider, the data collection of one tree is completed within minutes. Secondly, the collected B-scan is denoised through a signal processing framework proposed by leveraging free-space response removal, a column-connection cluster (C3)-based zero-gating algorithm, and finite impulse response (FIR) filtering [Fig. 1(a)]. The proposed signal processing framework removes the strong clutter due to reflection from the air-bark interface and makes the defect signature apparent in the B-scan. By doing so, it can achieve a 30 dB and a 22 dB increase in the signal-to-clutter and noise ratio (SCNR) of the measurement data of tree trunk samples and living trees, respectively. Finally, a novel multilevel feature fusion neural network (MLFF-Net) is proposed to detect defects in real time [Fig. 1(a)]. The proposed network extracts and fuses defects' signatures of different levels, evaluates the contribution of the fused feature maps in both channel and spatial dimensions, and classifies whether the tree trunk has or does not have a defect inside. The proposed network has achieved 96% classification accuracy for B-scans of the fresh-cut tree trunks.

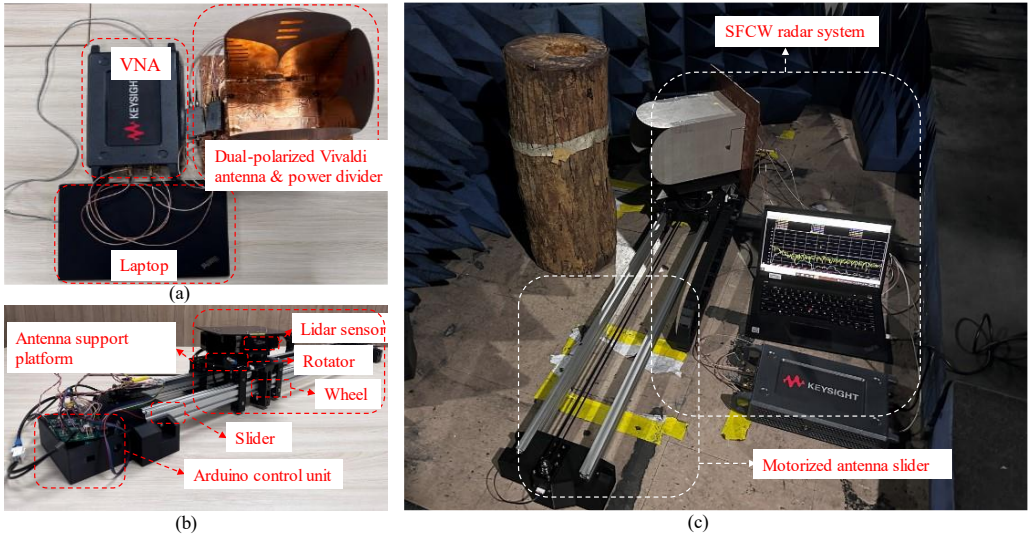

Fig. 2. (a) SFCW stand-off radar system. (b) Motorized antenna slider. (c) Scenario of scanning a tree trunk sample with SFCW system.

The main contributions of this work are as follows: 1) To the best of our knowledge, the proposed stand-off radar scheme is the first NDT technique, which allows rapid and contactless detection of trees' internal defects while operated on a straight scanning trajectory. The proposed measurement configuration completes the scanning of one living tree trunk within minutes without contacting the tree, which is more practical and less time-consuming compared to the existing circular scanning configurations. 2) The proposed signal processing framework is unique for denoising the noisy B-scans of real tree trunks. It successfully removes the clutter and reflections from the air-bark interface and reveals the features of tree defects in the B-scan with frequency weighting. 3) A novel MLFF-Net is specifically developed for detecting tree defects. (i) The multilevel features of defects are extracted and fused by the cascaded residual learning blocks (ResBlocks), and the feature fusion module, respectively. The fused feature maps naturally include a richer representation and better discrimination between different classes, especially for detecting diverse defects with various signatures from B-scans. (ii) A coordinate attention module (CAM) is employed to evaluate the contribution of the fused feature maps in both spatial and channel dimensions, which enables the network to automatically suppress the redundant information and emphasize the significant parts of the fused feature maps when predicting the existence of defects with various parameters. Tests with real measurement data show that the developed MLFF-Net outperforms the existing popular DL-based classifiers for tree defect detection. It should be noted that the earlier efforts on the development of the radar scheme were explained in [35], which briefly presented the signal processing framework. Compared to [35], this work presents a comprehensive study of the developed signal processing framework, which supplements the details of the methodology, a comparative study with existing signal processing techniques, and an extensive parametric study on the defect's size and type. In addition to that, the MLFF-Net is proposed for defect detection with extensive tests on different defective tree trunk samples. Last but not least, the work expounds the application of the proposed scheme to healthy and defective living trees, validating its capability in real-world scenarios.

The rest of this paper is organized as follows. The stand-off radar system and measurement configuration for scanning tree trunks are introduced in Section II. The developed signal processing framework to reveal the signatures of the defects in B-scans is presented in Section III. The MLFF-Net is described in Section IV. The applications of the signal processing framework on the measurement data of real tree trunk samples are provided in Section V. The accuracy of the MLFF-Net on the measurement dataset, along with the ablation study and the comparative study with the existing popular DL techniques, are expounded in Section VI. The results of the application of the developed stand-off radar scheme to defect detection in live trees are given in Section VII. Finally, the conclusions are drawn in Section VIII.

## II. STAND-OFF MEASUREMENT CONFIGURATION AND SFCW RADAR SYSTEM

In the measurement scenario presented in Fig. 1(b), a one-meter-long straight scanning trajectory is positioned adjacent to the tree trunk with the middle point 10 cm away from the bark. Once the measurement starts, the antenna is moved along the trajectory from one end to the other while its aperture always points towards the trunk center. Meanwhile, the radar system records the frequency response of the target tree and transforms it to a time-domain A-scan of each trace with every 2 cm movement of the antenna. By stacking a total of 51 recorded A-scans together, a raw B-scan is generated.

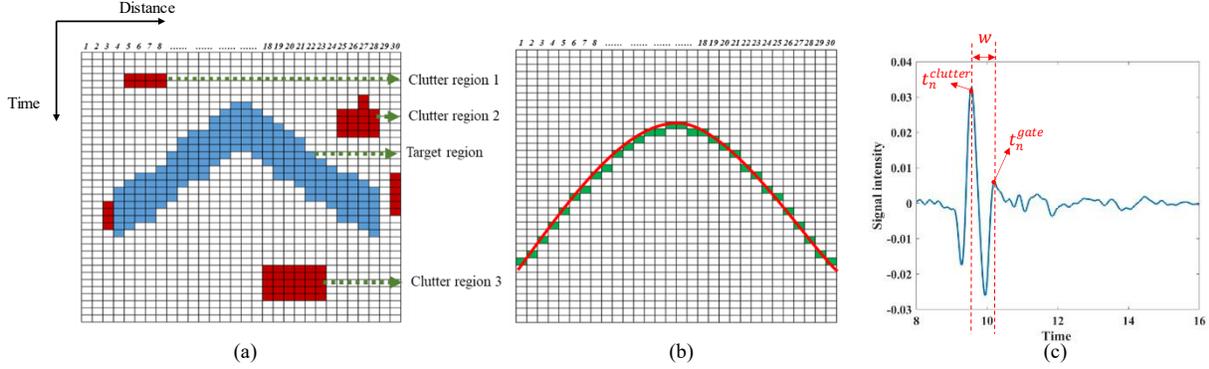

Fig. 3. (a) A binary image that contains ROIs. (b) Fitted hyperbola (red) and calculated clutter pixels (green) for all traces. (c) An example of the $n$th A-scan.

SFCW stand-off radar system [Fig. 2(a)] leverages a high-gain ultra-wideband dual-polarized Vivaldi antenna with narrow beamwidth [25]. The antenna consists of four elements. Each two parallel antenna elements are for one polarization and are connected with the same port of a vector network analyzer (Keysight VNA P5021A) through a power divider, which ensures uniform power delivery to each element. The laptop controls the VNA to record frequency responses from 0.5 to 4 GHz with 701 equally-spaced frequency points. The antenna is positioned on a motorized antenna slider, as shown in Fig. 2(b). This in-house slider was developed to realize the automation of the measurement process. Specifically, driven by the stepped motors inside the antenna support platform, the wheels under the platform move every 2 cm along the slider. At each scanning position, the lidar sensor detects the real-time distance of the tree trunk and calculates the correct rotation angle for the rotator to ensure that the antenna points toward the center of the tree trunk. To avoid measurement errors caused by mechanical vibrations of the platform during a real measurement [Fig. 2(c)], the system triggers the VNA to transmit and receive the signals many times at each scanning position. Once the received signal trace has less than 1% of the mean square error (MSE) compared to the previous trace at the same scanning position, the data is recorded, and the antenna is moved to the next scanning position. An Arduino control unit manages all the automation processes of the developed slider. All these components in the design of the radar system, including the specifically designed high-gain antenna, the selection of ultrawide low-frequency band, and keeping the aperture of the antenna facing the center of the tree trunk, maximize the energy transmitted from the radar system to the tree trunk and compensate for the power dissipation encountered during the penetration of the signal, which ensures the capability of defect detection in the stand-off scanning scheme.

## III. Signal Processing Framework

In the raw B-scan obtained via the stand-off scanning of the real tree trunk, the signatures due to the defects cannot be easily isolated from the clutter and noise in the measurement data, including the antenna's internal reflection, the clutter occurring at the air-bark interface, and environmental noise and clutter that are contained more in high-frequency components due to their weak penetration into the tree trunk. To address these issues, a signal processing framework consisting of three techniques shown in Fig. 1(a) is developed and applied to increase the SCNR and reveal the features of tree defects' reflections in the B-scan. First, free-space response removal, mentioned as a part of modelling the antenna-medium interaction mentioned in [36], [37] is performed to suppress the antenna's internal reflections due to the uneven impedance distribution between the feeding point and antenna aperture. Such internal reflections are estimated by holding the antenna towards the free space with no obstacle and measuring its return losses, which are further subtracted from each raw A-scan of the target tree to disclose the scattering information from the tree trunk in the B-scan. It should be noted that the free-space response removal cannot suppress the multiple signal reflections that occur between the antenna and the target when the antenna scans the tree trunk at a near distance. Then, a C3-based zero-gating algorithm [38] is introduced to automatically identify and remove strong hyperbolic clutter due to reflection from the air-bark interface. The details of the algorithm are provided next.

### A. Extractions of Regions of Interest (ROIs)

ROIs that contain high-intensity signals are separated from the background region. To achieve this, an adaptive threshold selection strategy mentioned in [38] is utilized to automatically determine the individual threshold value for each B-scan. In particular, the Sobel detector [39] is used to extract all boundaries between ROIs and the background. By averaging the signal intensities of selected boundaries, a proper threshold value is obtained to filter out the background with less intensity. The remaining B-scan containing only ROIs is then converted to a binary image for further clutter identification. For illustration purposes, an example of the binary image is shown in Fig. 3(a), where the blocks with colors indicate the retained ROIs, and each column refers to a single A-scan.

### B. Clustering Clutter Regions

Subsequently, the C3 algorithm is implemented and used to distinguish the target signal that represents the strong surface clutter (blue) from the high-intensity unwanted clutter (red) in Fig. 3(a). The fundamental unit in the algorithm is a segment that refers to a group of consecutive pixels in a column. Segments with sizes less than a threshold value $s$ are regarded as unwanted clutter. $s$ is set to be 5 for illustration purposes in Fig. 3(a), which classifies clutter region 1, clutter region 2, and the segment in column 3 successfully. Once segments of signals in each column are obtained, the clustering is processed from the first column to the last, searching for spatially connected segments in the images where "connected" means two segments in adjacent columns have pixels from the same row. The prior information, such as the target cluster's hyperbolic shape due to the straight scanning trajectory, helps filter out clutter region 3 while preserving the target region, which indicates the strong reflection from the bark in the B-scan at the end.

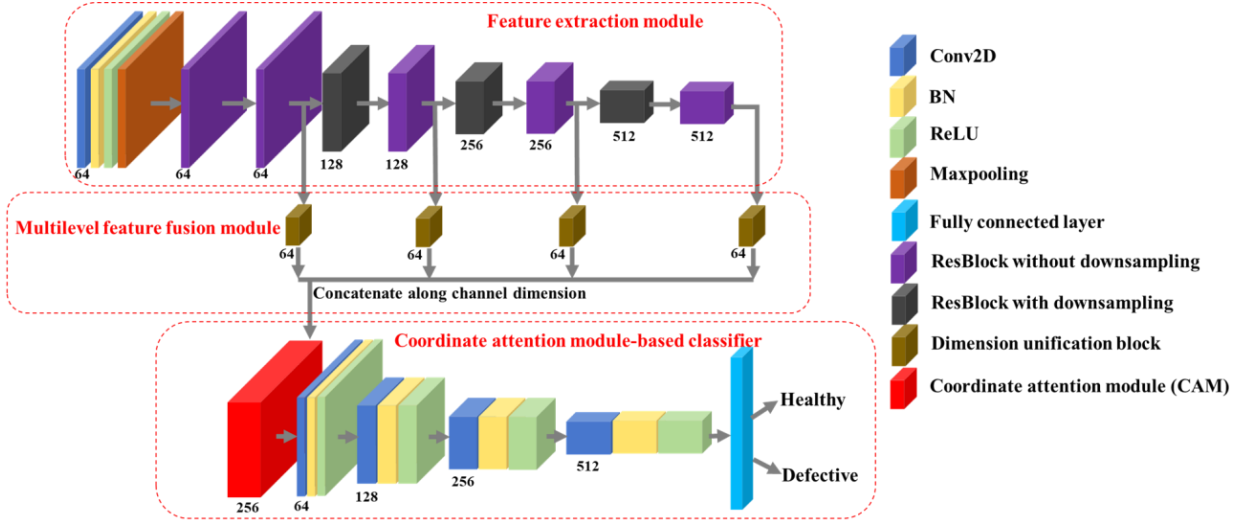

Fig. 4. Structure of MLFF-Net.

### C. Hyperbola Fitting

Due to the straight and linear scanning trajectory, the strong surface reflection from the bark is naturally a hyperbola in the B-scan. Despite some segments completing the hyperbolic shape in the target region (in columns 1, 2, 3, 29, and 30 of Fig.3(a)), they are removed after thresholding. To determine the row indices of the segments that are not clustered by the target region, a hyperbola fitting is performed by considering the middle pixels of the segments in the target region as

$$\frac{(n-d)^2}{a^2} + \frac{t_n^2}{b^2} = c,$$  (1)

where $n$ is the column index of the middle pixel of the segment in the target region, and $t_n$ is the row index in the $n$th column. The parameter set $(a, b, c, d)$ is obtained by minimizing MSE with the constraint of $10 < d < 20$, indicating that the vertex of the hyperbola is near the middle trace on the scanning trajectory.

### D. Zero-gating

Fig. 3(b) shows the fitted hyperbola (red) and the calculated clutter pixels (green) in all traces. Since the time location $t_n^{clutter}$ of the clutter pixel in trace $n$ usually represents the high-intensity signal around the maximum of the clutter pulse in Fig. 3(c), a compensation term $w$ that considers the width of the pulse ensures the complete removal of the clutter region. The zero-gating reference time $t_n^{gate}$ of trace $n$ is then calculated by

$$t_n^{gate} = t_n^{clutter} + w$$  (2)

Therefore, all the segments before $t_n^{gate}$ are set to zero to distinguish the strong surface clutter from the late-time signals that contain weak reflections from the tree's interior structures in the processed B-scan.

Finally, an FIR filter generated by the classical Kaiser window [40] is applied to the zero-gated B-scan to further improve the SCNR. The classical Kaiser window is preferable because its window shape can be conveniently controlled without changing the window length. Since environmental clutter and noise are contained more in high-frequency components due to their weak penetration into the tree trunk, a Kaiser window with a central frequency of 1 GHz is applied to the frequency samples within the band of [0.5, 4] GHz, obtained after Fourier transforming zero-gated A-scan. Such FIR filter application suppresses the noise and enhances the potential to detect defects in the processed B-scan.

## IV. MULTILEVEL FEATURE FUSION NEURAL NETWORK

Despite the applications of signal processing techniques, the signatures of defects with different positions, sizes, and types are diverse in terms of scales, locations, and shapes in the processed B-scan, hindering the detection of tree defects. Inspired by the Feature Pyramid Network [41], which extracts and fuses multilevel feature maps to identify various objects with independent scales, an MLFF-Net is developed to detect the features of various tree defects. Fig. 4 shows the overall structure of the proposed MLFF-Net consisting of three main components: A feature extraction module, a multilevel feature fusion module, and a CAM-based classifier, details of which are provided below.

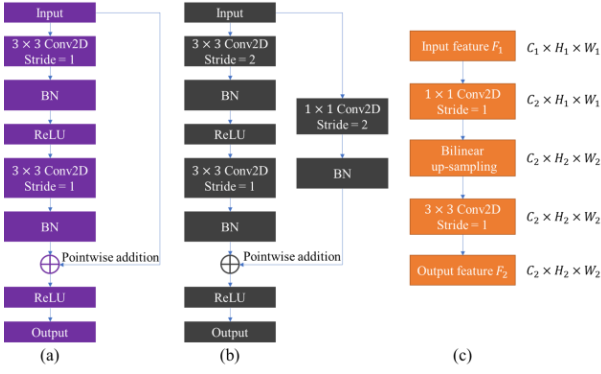

Fig. 5. Structures of: (a) ResBlock without down-sampling, (b) ResBlock with down-sampling, and (c) dimension unification block.

### A. Feature Extraction

The feature extraction module takes the processed B-scans as the inputs, interprets the diverse characteristics of features due to various tree defects, and extracts their features into multiple levels of feature maps. Specifically, the input is first passed through a preprocessing block, consisting of a $7 \times 7$ convolution (Conv2D) layer with a stride size of $2 \times 2$, a batch normalization (BN) layer, the rectified linear unit (ReLU) activation function, and a $2 \times 2$ max pooling layer (Maxpooling). Such a large convolutional layer not only captures more global information but also reduces the volume size of the output feature maps to save the computation cost.

Then, eight successive ResBlocks [42], including blocks with and without down-sampling operations, are applied to extract various features of tree defects from low level to high level accordingly. A classical ResBlock is shown in Fig. 5(a) where the residual learning is achieved by the shortcut connection between the input and the output of the second BN layer to avoid degradation issues in the learning process. The down-sampling in Fig. 5(b) is realized by adjusting the first Conv2D layer in the main path and the $1 \times 1$ Conv2D layer in the shortcut connection path, strides of which are set to be 2 to reduce the dimension of the output feature map. The channel numbers of eight ResBlocks are set as [64, 64, 128, 128, 256, 256, 512, 512], in which the repetitive setting of the channel number guarantees thorough learning after each down-sampling operation.

### B. Multilevel Feature Fusion

Different from conventional convolutional neural networks, such as ResNet [42], DenseNet [43], and VGG [44], which utilize the highest-level feature map obtained from the end of the feature extraction module for the classification task, the MLFF-Net fuses feature maps obtained by the Resblocks from different levels. The fused feature map naturally retains more comprehensive information on the diverse features of defects, which is significant for detecting defects with various parameters from the B-scan. To realize the fusion of multilevel feature maps with different sizes, each feature map is converted to the same size by the dimension unification block [41] which is shown in Fig. 5(c). Considering the input and output feature maps are with the sizes of $C_1 \times H_1 \times W_1$ and $C_2 \times H_2 \times W_2$, respectively, a $1 \times 1$ Conv2D layer with filter number $C_2$ first revises the channel number of the feature map in the channel dimension, and then an up-sampling operation through the "bilinear" interpolation unifies the size in the spatial dimension. Lastly, a $3 \times 3$ Conv2D layer suppresses the redundant information induced by the interpolation. Once the feature maps of all levels are unified to the same size with a channel number of 64 in this study, feature fusion is completed by concatenation in the channel dimension, resulting in a 256-channel fused feature map with a rich diversity of features.

### C. Coordinate Attention Module-based Classifier

Besides including different levels of representations of various defect signatures, the fused feature maps contain features of interference signals and redundant information that impair the accuracy of defect detection. As a result, a CAM [45] is introduced to automatically encode the multilevel features based on their contributions to the correct predictions in both spatial and channel dimensions before feeding them to a classifier. Such contributions are known as "attention" which allows the model to "pay attention" to certain parts of the fused feature maps and to give them more weight when predicting the existence of defects with diverse signatures. During the evaluations of the attention of the fused feature maps, the important features representing the defects are emphasized, while the features of interference signals are suppressed, which improves the capability of the following classifier in detecting various defect signatures. The schematic of the CAM is presented in Fig. 6, and the encoding process is described as follows.

Given an input feature map $X \in R^{C \times H \times W}$ [Fig. 6], where $C$, $H$, and $W$ are the channel number, the height, and the width, respectively. Two one-dimension average pooling operations first encode features of each channel along the width and height coordinate separately, generating two corresponding channel descriptors $z^h$ and $z^w$ [Fig. 6]. The $c$ th channel of the descriptors at a given height and width is written as:

$$z_c^h(h) = \frac{1}{W} \sum_{0 \le i < W} X_c(h, i) \tag{3}$$

$$z_c^w(w) = \frac{1}{H} \sum_{0 \le j < H} X_c(j, w) \tag{4}$$

Such transformations allow the CAM to capture long-range interactions along one direction while preserving accurate features along the other, which is beneficial for locating the multilevel characteristics of the defects. Subsequently, the direction-aware descriptors are concatenated and passed to a shared Conv2D block $L_1$ yielding an intermediate feature map $IF \in R^{C/r \times (W+H)}$:

$$IF = L_1([z^h, z^w]) \tag{5}$$

where $[\cdot, \cdot]$ denotes the concatenation operation along the spatial dimension. $r$ is the reduction ratio in the $L_1$ block to save the computation cost. The attention weight of two directions is then encoded independently by splitting the $IF$ into two separate feature map $IF^h \in R^{C/r \times H}$ and $IF^w \in R^{C/r \times W}$. Two additional Conv2D blocks $L_h$ and $L_w$ are deployed to produce two separate attention weights $k^h \in R^{C \times H}$ and $k^w \in R^{C \times H}$ expressed as

$$k^h = L_h(IF^h) \qquad (6)$$
$$k^w = L_w(IF^w) \qquad (7)$$

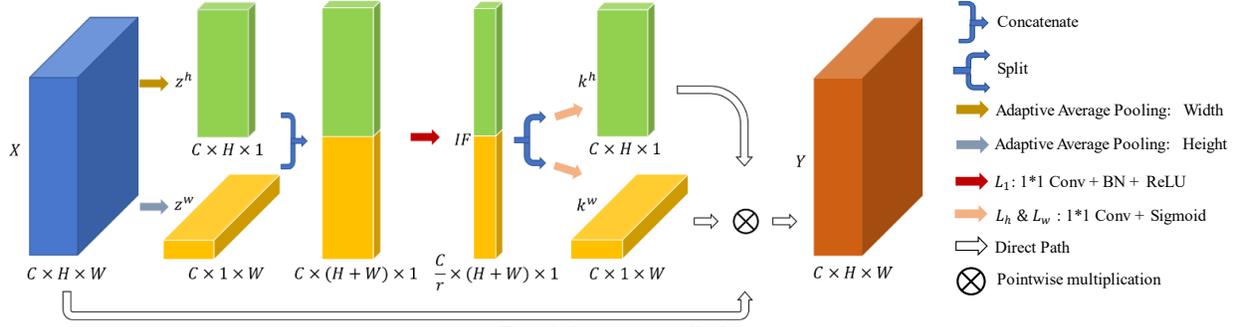

Fig. 6. Structure of CAM.

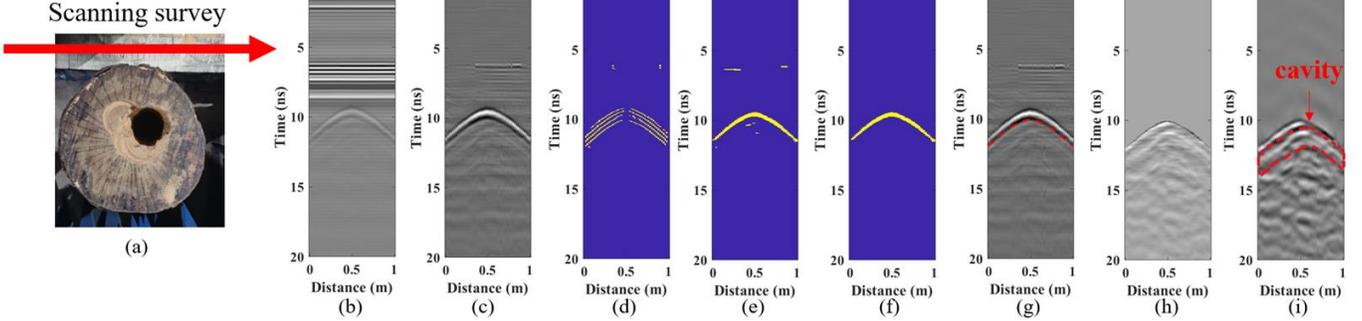

Fig. 7. (a) Tree trunk sample and (b) raw B-scan. The output of the steps of signal processing framework: after (c) applying free-space response removal, (d) executing edge detection, (e) finding ROIs, (f) determining clutter via C3 algorithm, (g) applying zero-gating time index, (h) removing the clutter from B-scan, and (i) applying FIR filtering.

TABLE I
SCNR OF B-SCANS AT DIFFERENT PROCESSING STAGES

|  | Raw data | Free space response removal | C3-based zero-gating | FIR filtering |
|---|---|---|---|---|
| SCNR (dB) | -13.91 | 3.55 | 15.02 | 16.41 |

The sigmoid function inside the block interprets the values of attention weights into an interval of [0, 1]. Consequently, the automatically weighted feature map $Y \in R^{C \times H \times W}$, which encodes the inter-channel relationship as well as the spatial dependence along the corresponding direction in the intermediate feature maps, is simplified as

$$Y_c(i,j) = X_c(i,j) \times k_c^h(i) \times k_c^w(j) \qquad (8)$$

where $c$ represents the index of the channel dimension. By employing the CAM, the multilevel fused feature map is naturally well-weighted, so contributions from different levels are adaptively adjusted in terms of sizes, positions, and types of defects. It should be noted that the complete structure of the CAM in Fig. 6 is given as a single red block located at the beginning of the "Coordinate attention module-based classifier" part in Fig. 4. Finally, the attention-weighted feature map with the channel number of 256 is forwarded into a conventional classifier that consists of four successive Conv2D blocks with the set of channel numbers to be [64, 128, 256, 512], and one fully connected layer followed by the sigmoid function to predict the existence of the defects. Each Conv2D block includes a Conv2D layer with a stride size of $2 \times 2$, a BN layer, and the ReLU activation function.

## V. APPLICATION OF SIGNAL PROCESSING FRAMEWORK TO THE MEASUREMENT DATA OF REAL TREE TRUNKS

In this section, experimental data of a real tree trunk sample collected by the measurement setup explained in Section II is used to test the signal processing framework introduced in Section III. The tree trunk sample with a diameter of 30 cm and a height of 50 cm is a section of a fresh-cut Angsana (*Pterocarpus indicus*) trunk sample, a common roadside tree species found in Singapore.

### A. Enhancement of Defect Signature

To study the features of defects in the B-scan, a cylindrical hole with a diameter of 6 cm is drilled into the trunk sample [Fig. 7(a)]. The measurement trajectory is on the top side of the tree trunk model with the scanning direction from the left to the right side [Fig. 7(a)]. Fig. 7(b)-(i) shows the application of the signal processing framework to the collected raw B-scan. It should be noted that all the B-scans are time-referenced to the power divider where we calibrated the radar system. In Fig. 7(b), the signatures from the tree trunk are almost hidden by the antenna's internal reflections. After free-space response removal, the hyperbola-shaped clutter due to reflection from the air-bark interface is revealed [Fig. 7(c)]. This clutter hinders seeing the signatures of defects in the B-scan. After applying the edge detection [Fig. 7(d)] and the adaptive threshold selection, the binary image that includes all candidates of pixels for clustering is obtained as in Fig. 7(e). In addition to the hyperbola-shaped curves due to reflection from the air-bark interface, some pixels of unwanted clutter appear above and below the curves in the B-scan [Fig. 7(e)]. By setting proper threshold values, applying clustering rules, and performing hyperbola fitting, the unwanted clutter is removed, and the strong surface clutter reflected from the air-bark interface is revealed as in Fig. 7(f). The zero-gating reference time (red dashed line in Fig. 7(g)) is set by adding the pulse width to the middle position of the clutter region to completely eliminate the air-bark clutter as in Fig. 7(h). After applying an FIR filter generated by a Kaiser window with a central frequency of 1 GHz, the signature due to the cavity's reflection, an internal hyperbolic curve, is distinguished easily from other noise patterns [Fig. 7(i)]. Since the cavity is located off-center, the apex of the corresponding hyperbolic curve shifts to the right side of the B-scan, as in Fig. 7(i). To quantitively illustrate the effectiveness of the signal processing framework, the SCNR [46] of the B-scan is calculated at each stage of the framework [Table I]. It is apparent from Table I that the SCNR is significantly improved by 17.46 dB and 11.47 dB after removal of free-space response and air-bark clutter, respectively. After 1.39 dB SCNR improvement by FIR filtering, the framework achieves more than 30 dB SCNR improvement overall.

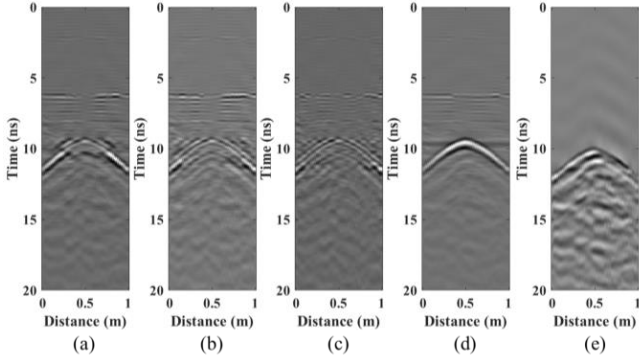

Fig. 8. Processed B-scan by: SVD with (a) 2, (b) 4, and (c) 6 singular values and vectors being removed, (d) background removal, and (e) C3-based zero-gating algorithm.

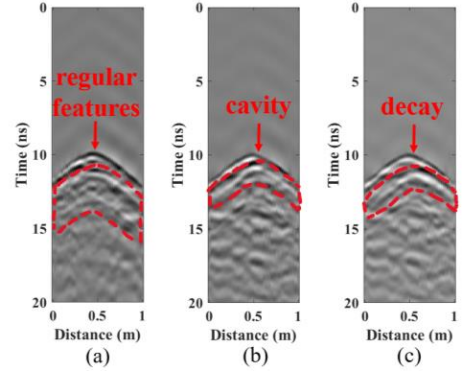

Fig. 9. Processed B-scan of tree trunk (a) without any defect, (b) with cavity, and (c) with decay.

### B. Comparative Study with Existing Clutter Removal Techniques

The effectiveness of the C3-based zero-gating algorithm is demonstrated by comparing its performance with conventional clutter removal techniques, including background removal [9], and SVD [11]. For a fair comparison, different numbers of dominant singular values and vectors are removed when the B-scan is processed by SVD. All these techniques are applied to the same B-scan after removing the antenna's internal reflection [Fig. 7(c)] and the results are shown in Fig. 8. Although removing an increasing number of singular values through SVD weakens the intensity of the strong surface clutter [Fig. 8(a)-(c)], the hyperbolic defect is distorted as well, resulting in indistinguishable patterns of the defect in the processed B-scan.

Moreover, the surface clutter is not affected by the background removal in Fig. 8(d) because the clutter is not time constant response due to the straight scanning trajectory. Compared to the other two techniques, the developed C3-based zero-gating algorithm removes the strong surface clutter entirely while retaining the integrity of the defect signatures in the processed B-scan [Fig. 8(e)], making the patterns of defects easily distinguishable.

### C. Study on Signatures of Defects with Different Types

The proposed signal processing framework is applied to the raw B-scans of the tree trunks with different defects (cavity and decay) and without any defects to investigate the signatures of defects. To this end, the B-scan of the tree trunk with the cavity considered in the previous part is used. In addition, the B-scan of the tree without any defect is obtained before drilling the cavity inside the same trunk in Fig. 9(a). The B-scan of the trunk with decay is simulated by filling the cavity of the same tree trunk in Fig. 7(a) with paper mâché. This decay replicate is soaked into the water before measurement to mimic the high-water contents of real wood decay.

The B-scans processed by the signal processing framework are shown in Fig. 9. Despite some noise patterns induced due to inhomogeneous wood material, the dominant features in the B-scan of the tree trunk without any defect [Fig. 9(a)] are quite regular: Two hyperbolic curves are parallel to the near-end zero-gating boundary while their left wings start at 12 ns and 14 ns (at 0m scanning distance), respectively [Fig. 9(a)]. The first hyperbolic curve is due to the internal reflection from the concentric layer of the tree trunk, while the second hyperbola corresponds to the reflection from the far-end surface. Such a regular pattern is significantly distorted by introducing the cavity, which changes the overall structure of the tree trunk, resulting in an abnormal pattern [Fig. 9(b)]. Since the cavity is located at the trunk's upper half and near the scanning trajectory, the cavity signature is closer to the reflection from the air-bark interface. Due to the high permittivity and conductivity of the wet tissue, the intensity of the signal that represents the decay signature in Fig. 9(c) is much stronger, especially at scanning distances larger than 0.5 m. Moreover, the apex of the hyperbolic signature shifts more to the right side because of the off-centered position of the decay. Although signatures of defects are visible in both Fig. 9(b) and Fig. 9(c), their patterns vary with respect to their sizes, locations, and types, therefore challenging the tree defect detection from a noisy pattern of the processed B-scan, which again emphasizes the necessity of the proposed MLFF-Net.

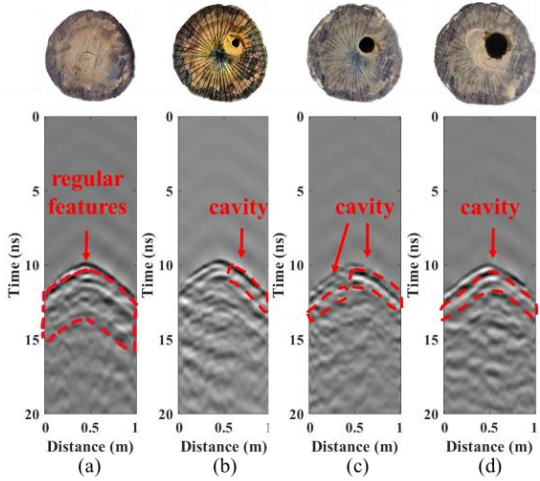

Fig. 10. Processed B-scan with the diameter of the cavity to be (a) 0 cm, (b) 2 cm, (c) 4 cm, and (d) 6 cm.

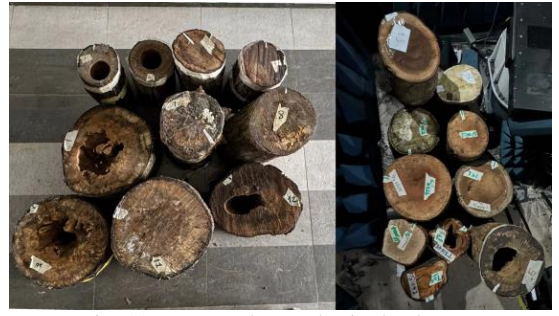

Fig. 11. Tree trunk samples in the dataset.

### D. Study on Signatures of Defects with Different Sizes

To study the variation of the signatures of defects with different sizes in the processed B-scans, a set of measurements is conducted on the Angsana tree trunk sample in which cylindrical holes with increasing diameters are drilled into the same position. The drilled tree trunk samples and corresponding processed B-scans are shown in Fig. 10. The scanning trajectory is on the top side of the tree trunk sample with the direction from left to right. For convenient comparison, Fig. 10(a) and Fig. 10(d) are the same as Fig. 9(a) and Fig. 9(b), respectively.

Based on the analysis in Section V.C, the signature of late arrival signals after the clutter due to air-bark surface reflection is overall regular in the B-scan of tree trunk without defect [Fig. 10(a)]. The influence of the diameter of the hole on the processed B-scan can be investigated in three cases. First, when the diameter of the hole is as small as 2 cm, the continuous pattern of the internal reflection in the measurement of the healthy sample is distorted at the scanning distance of less than 0.5 m. Meanwhile, an additional continuous response appears in the marked region of Fig. 10(b), which implies the presence of discontinuity inside the tree trunk sample. Besides that, the position of the signatures also indicates that the cavity should appear in the top-right region of the sample. The second case is an intermediate stage with the diameter of the cavity being 4 cm [Fig. 10(c)]. The signal reflected from the cavity captured at the scanning distance larger than 0.5 m becomes much stronger due to the closer distance between the cavity and the near-end surface. Moreover, the signature of the cavity captured at a scanning distance of less than 0.5 m appears to be visible in the marked region on the left. In the third case, when the diameter of the cavity is extended to 6 cm, the signature induced by the cavity is totally visible and distinguishable from the processed B-scan in Fig. 10(d). The blurred signature in the left-marked box [Fig. 10(c)] is now totally connected, indicating that the reflections received at the first 50 cm of the scanning trajectory are strong enough to be visible. Moreover, the signatures in both marked regions [Fig. 10(c)] are completely merged, forming a continuous hyperbolic signature induced by the cavity. To conclude, reflections from defects with small sizes can distort the pattern of internal reflection and generate abnormal signatures in the processed B-scan. As the defects become larger, the magnitudes of the corresponding reflection signals are large enough to be visible and distinguishable from the background pattern, while only a partial of the signatures are observed when the size of the defect is at the intermediate stage.

## VI. APPLICATION OF MLFF-NET WITH MEASUREMENT DATASET

### A. Dataset Preparation

The proposed MLFF-Net is trained and tested with B-scans of real tree trunk samples collected via the proposed stand-off radar system. The dataset consists of two classes, including the B-scans of the trunks without defects, labeled as "healthy", and B-scans of the trunks with defects, labeled as "defective", respectively. To generate the dataset, twenty tree trunk samples [Fig. 11] with convex shapes are obtained from live Angsana trees, including ten healthy tree trunks and ten defective tree trunks. The diameters and heights of these tree trunks vary within [20, 45] cm and [35, 120] cm, respectively. The defective samples contain tree trunks with drilled defects and naturally grown internal defects. The closest distance between defects and barks in the dataset is 4 cm. The geometrical properties of the collected samples could cover most of the Angsana trees in Singapore. To avoid the large reduction of the moisture content of the freshly cut tree trunk samples in the laboratory environment, the samples are scanned shortly after collection. Since characteristics of defects in the B-scan vary with respect to the locations of defects, such diversity is included in the dataset by conducting measurements on each sample from various rotation angles from $0°$ to $350°$ with a stepping angle of $10°$. To consider the influence of different scanning directions on the signatures in the B-scan, two sets of measurements, including scanning starting from both ends of the straight trajectory, are conducted at each rotation angle of the tree trunks. In total, 1440 measured B-scans are collected, and each category contains 720 measured B-scans.

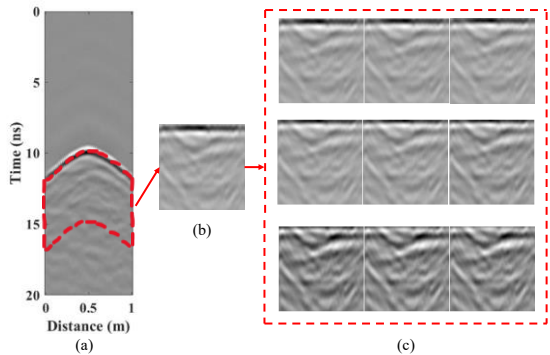

Fig. 12. (a) Processed B-scan, (b) reshaped B-scan, and (c) additional B-scans with various values of the compensation term.

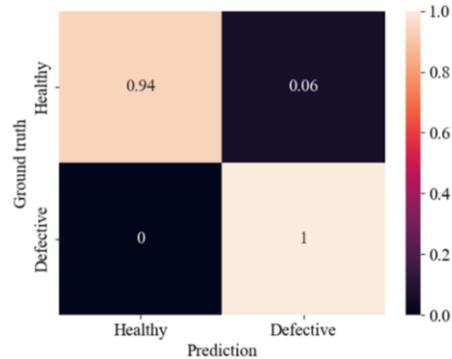

Fig. 13. Confusion matrix.

TABLE II
ABLATION STUDY

| Model | A | B | C |
|---|---|---|---|
| Feature fusion | ✗ | ✓ | ✓ |
| CAM | ✓ | ✗ | ✓ |
| Acc. (%) | 87.01 | 90.35 | **91.38** |
| Prec. (%) | 87.91 | 92.27 | **92.73** |
| Rec. (%) | 87.01 | 90.35 | **91.38** |
| F1 score (%) | 86.81 | 90.03 | **91.15** |

TABLE III
COMPARISON OF PERFORMANCE OF MLFF-NET WITH DIFFERENT
ATTENTION MODULES

| Network | Acc. (%) | Prec. (%) | Rec. (%) | F1 score (%) |
|---|---|---|---|---|
| MLFF+SEM | 89.51 | 91.14 | 89.51 | 89.12 |
| MLFF+CBAM | 89.31 | 91.63 | 89.31 | 88.85 |
| **MLFF+CAM** | **91.38** | **92.73** | **91.38** | **91.15** |

After obtaining a processed B-scan shown in Fig. 12(a), three additional adjustments are applied to generate diverse input for the MLFF-Net. First, the hyperbolic zero-gating curve determined by the C3 algorithm is straightened and regarded as the zero-time in each trace so that the redundant space before the first reflection from the bark is fully eliminated. Second, the time duration in the processed B-scan image is limited to 5 ns by considering the size and equivalent permittivity of the measured samples, which helps to isolate the defect features if they are present. The cropped B-scan image is resized to the resolution of $128 \times 128$ shown in Fig. 12(b). Third, since the normalized magnitudes of the signals reflected from the defects are sensitive to the zero-gating indices, which influence the range of the signal intensity in the B-scan, the slight variation of the zero-gating compensation term $w$ in (2) can generate visually different features of defects. To address this issue, an additional compensation term $w_n = 0.03n$ ns for $n \in Z, 1 \le n \le 9$, is added to expression in (2) to generate 9 additional B-scans [Fig. 12(c)] with different zero-gating indices. The generated B-scans are stacked with the original B-scan with $w_0 = 0$ ns (in [Fig. 12(b)]) and inputted into the network from 10 channels to ensure consistent predictions.

### B. Implementation Details of MLFF-Net

After obtaining the dataset, five-fold cross-validation is applied to avoid bias in the performance metrics due to the limited dataset size. To be specific, 20 tree samples are randomly divided into five folds, each fold contains 4 tree samples, including 2 healthy tree samples and 2 tree samples with defects, resulting in a total number of 288 B-scans. Alternatively, the model is fitted by using four of the five folds as the training dataset while being validated by the remaining fold, resulting in five well-trained models with different combinations of training and testing data. During the training of the proposed MLFF-Net, the cross-entropy loss is used, and the network's parameters are optimized by the Adam optimizer. The network is trained on an NVIDIA RTX 6000 GPU for 100 epochs with a learning rate of 0.0005 and a batch size of 64, during which the model with the highest testing accuracy is saved. Apart from the testing accuracy, additional metrics, including the precision, recall, and F1 score are computed to quantitatively evaluate the performance of the MLFF-Net. Considered a classification problem with $N$ categories, the precision, and recall for the $i$th class $Pr_i$ and $Re_i$ are calculated by

$$Pr_i = TP_i / (TP_i + FP_i) \qquad (9)$$
$$Re_i = TP_i / (TP_i + FN_i), \qquad (10)$$

where $TP_i$, $FP_i$, and $FN_i$ represent the numbers of true positive, false positive, and false negative for Class $i$, respectively. The F1 score of Class $i$, $F1_i$, is expressed as:

$$F1_i = 2 \times Pr_i \times Re_i / (Pr_i + Re_i) \qquad (11)$$

After computing all these metrics for all classes, their macro averages are computed. For example, the macro average F1 score, $F1$, is computed via $F1 = (1/N) \sum_{i=1}^{N} F1_i$, where the importance of each class is regarded as the same.

### C. Performance of MLFF-Net

The best MLFF-Net model achieves an accuracy of 96.88%, a precision of 97.06%, a recall of 96.88%, and an F1 score of 96.87% with the measurement data of the real tree trunk samples. When five-fold cross-validation is applied, the average accuracy, precision, recall, and F1 score of all trained five models are obtained as 91.38%, 92.73%, 91.38%, and 91.15%, respectively. The confusion matrix obtained by the best MLFF-Net model is provided in Fig. 13. The model predicts defect class with 100% accuracy while misclassifying 6% of the samples in healthy class as in defective class. Given that the network is trained with a very limited dataset (1152 B-scans), it can be foreseen that the performance of the model could be enhanced by increasing the size of the training dataset. In general, all these high-performance metrics indicate the robustness of the proposed MLFF-Net in accurately detecting the defects from provided B-scans.

TABLE IV
COMPARISON OF THE PERFORMANCES OF THE PROPOSED MLFF-NET AND OTHER DL TECHNIQUES

| Network | Acc. (%) | Prec. (%) | Rec. (%) | F1 score (%) | Memory (GB) |
|---------|----------|-----------|----------|--------------|-------------|
| VGG-16 | 82.71 | 84.49 | 82.71 | 84.49 | 11.65 |
| ResNet-18 | 88.96 | 90.16 | 88.96 | 88.53 | 1.09 |
| DenseNet | 87.50 | 90.40 | 87.50 | 86.73 | 1.64 |
| VIT-16 | 52.29 | 28.91 | 43.82 | 33.68 | 24.35 |
| CNN-MLP | 84.93 | 87.27 | 84.93 | 84.66 | 6.10 |
| **MLFF-Net** | **91.38** | **92.73** | **91.38** | **91.15** | **1.38** |

## D. Ablation Study

An ablation study refers to an experiment in which certain modules of a deep learning model are systematically removed to analyze their particular contributions to the model's performance. To demonstrate the necessity of two main components of the MLFF-Net, including the feature fusion module and CAM, two ablated models are retrained accordingly with all components, except for the ablated part, being kept the same. To avoid selection bias due to the limited size of the dataset, five-fold cross-validation is again applied for reliable evaluation of the models in this and next subsections. Moreover, the reported results are averages of the performance metrics of five trained networks.

Table II lists the performance of the ablated models on the same dataset where the proposed MLFF-Net is denoted as Model C. In Model A, the multilevel feature fusion module (in Fig. 4) is removed and the features extracted from the last ResBlock are directly imported to the CAM-based classifier. As presented in Table II, nearly all metrics of Model A are around 4% lower than those of the proposed Model C, indicating the substantial enhancement of the detection capability by embedding multilevel feature fusion in the MLFF-Net. In Model B, the CAM (in Fig. 4) is removed from the classifier to evaluate the contribution of the attention mechanism to the MLFF-Net. The comparison between the performance metrics of Model B and Model C verifies the significance of the contribution of weighting fused features before inputting to the classifier. To further investigate the effectiveness of the CAM in the MLFF-Net, other well-recognized attention modules, including the squeeze and excitation module (SEM) [47] and the convolutional block attention module (CBAM) [48], are embedded into the network by replacing CAM. Their performances are compared with those of CAM [Table III]. Compared to the structure of the CAM, the SEM only considers the channel dimensional dependency, whereas the CBAM encodes the spatial features into a solo value, which ignores the spatial correlations along both spatial dimensions of a feature map. Therefore, it is not surprising that CAM outperforms all other attention modules and the deployment of the CAM in MLFF-Net yields the best performance metrics. Overall, the implementation of all main components of the designed MLFF-Net contributes to the highest performance, demonstrating the significance of fusing multilevel feature maps and evaluating the corresponding attention for detecting diverse defect signatures in the measurement data.

## E. Comparative Study with Existing DL Techniques

The effectiveness of the proposed MLFF-Net is proved by comparing its performance with those of convolutional neural network (CNN)-based DL techniques, including the VGG [44], the ResNet [42], and the DenseNet [43], and also a transformer-based method, ViT [49]. Furthermore, the CNN-MLP model [34], validated on the synthetic dataset, is also used for comparison. To have a fair comparison, the number of the input channels of all models is revised to 10 for using the same dataset. Training and testing settings in all networks are kept the same.

Table IV compares the performances of the proposed MLFF-Net and the existing DL methods in terms of accuracy and memory requirements. It is clear in Table IV that ViT-16 performs the worst due to the large dataset requirement of transformer-based techniques. Among CNN-based classifiers, VGG-16 performs the worst while it requires the largest memory. Just like VGG-16, CNN-MLP exhibits a poor performance compared to other CNN-based classifiers. In contrast, DenseNet achieves high accuracy with a relatively low memory requirement, while the ResNet-18 is the most accurate one with a lower memory requirement compared to DenseNet. Such a comparison indicates that capturing features by completely cascading the convolutional layers in the VGG-16 (or in CNN-MLP) is not effective for the measurement data of real tree trunk samples. Instead, alternative ways of transmitting features, such as the shortcut connections in the ResNet-18 and the bypass connections in the DenseNet, largely increase the diversity of the feature, therefore alleviating the degradation of the model with a small dataset. Although the proposed MLFF-Net utilizes the same feature extraction block as that in ResNet-18, its accuracy is even better due to the contribution of the feature fusion module as well as the CAM.

## VII. APPLICATION OF THE PROPOSED STAND-OFF RADAR SCHEME TO THE DETECTION OF THE DEFECT IN LIVE TREES

### A. Tests on Living Tree with Defect

The performance of the proposed stand-off radar scheme is examined via its application to the detection of a cavity inside a live tree [Fig. 14.(a)]. A live *Araucaria cunninghamii* with a diameter of 46 cm is selected for the measurement as it has an internal cavity at 1-meter height, determined by another testing technique. Using the same measurement configuration (explained in Section II), four raw B-scans are obtained while the mid-point of the slider is positioned near the four different sides of the tree [Fig. 14.(a)]. The scanning is performed from right to left while the antenna is facing towards the tree. After the measurements, the tree is cut down and the cross-section of the scanned portion is photographed [Fig. 14.(b)], where a cavity with an irregular shape is observed.

The proposed signal processing framework is used to process four raw B-scans and reveal the signatures of defects. In Fig. 15(a) and (b), the pulses reflected from the cavity are hyperbolas with 180 degrees of phase-shifted, which reverses the pulse reflected from the bark due to the difference in the refractive indices of the two media. The relatively smooth outer shapes of the cavity seen from the scans from Position 1 and 2 produce regular and continuous hyperbolic features of defects in the B-scan. Although such characteristics of defects are degraded in the B-scans at Position 3 and 4 because of the corresponding tortuous outlines of the cavity, the existence of defects could be verified by visible abnormal defect signatures in Fig. 15(c) and (d). Moreover, as indicated in Table V, the signal processing framework improves the SCNR of the B-scans of the live tree around 22-25 dB.

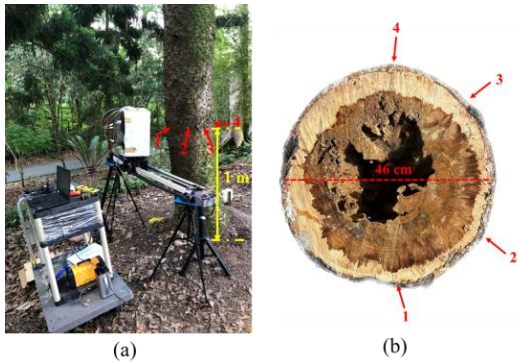

Fig. 14. (a) On-field measurement setup. (b) Photo of the cross-section of the scanned portion of the defective tree trunk (obtained after being cut).

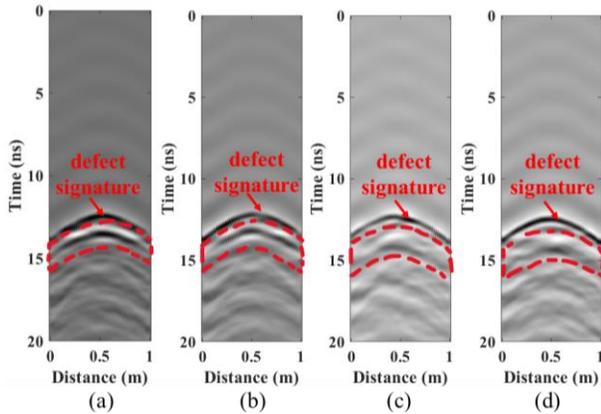

Fig. 15. Processed B-scan of measurements conducted at: (a) Position 1, (b) Position 2, (c) Position 3, and (d) Position 4.

The processed B-scans are then imported to the well-trained MLFF-Net for the rapid defect detection and health diagnosis of the target tree. As can be seen in Table V, the network recognizes the diverse signature of the internal cavity from 3 of 4 processed B-scans successfully within seconds, achieving an accuracy of 75%.

### B. Tests of False Alarm Rate on Healthy Living Trees

False alarm rate is significant to avoid the incorrect identification of healthy living trees as trees with defects. To evaluate the false alarm rate of the proposed scheme, four healthy living trees with diameters ranging from 25 to 45 cm are selected by the arborist [Fig. 16]. Then scanning of each tree trunk is conducted from four different sides by the developed stand-off radar system, resulting in a total amount of 16 B-scans. After the measurements, all trees are further verified to contain no defects at the scanning height by the invasive but reliable drilling resistance method. The collected data is processed by the developed signal processing framework accordingly. For illustration purposes, one selected processed B-scan from each tree dataset is presented in Fig. 16. It can be observed that the signatures of the late arrival signals after the surface clutter are relatively regular in the processed B-scans of healthy living trees.

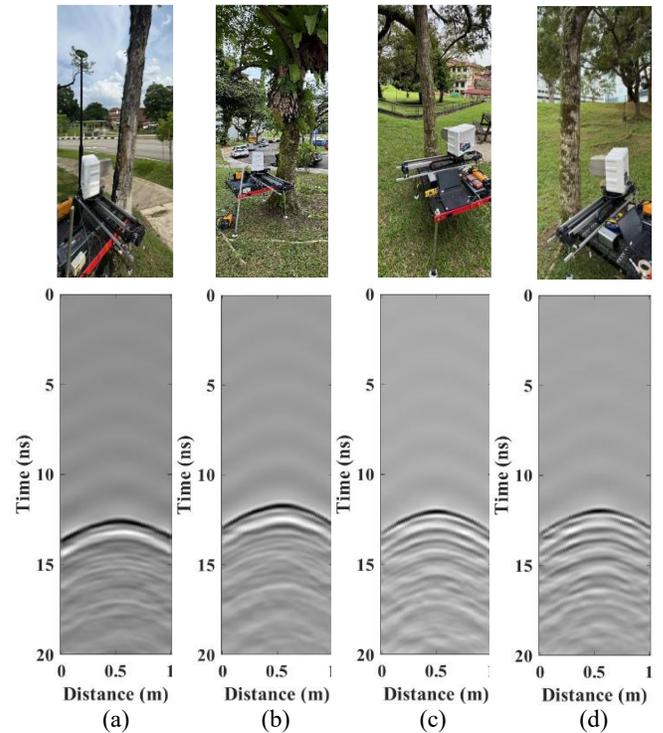

Fig. 16. Measurement scenarios and processed B-scans of: (a) Tree 1, (b) Tree 2, (c) Tree 3, and (d) Tree 4.

TABLE V
SCNR ENHANCEMENT AND PREDICTION BY PROPOSED SYSTEM

| Scanning position | 1 | 2 | 3 | 4 |
|---|---|---|---|---|
| SCNR improvement (dB) | 25.41 | 25.96 | 22.97 | 25.93 |
| Prediction by MLFF-Net | Defective | Healthy | Defective | Defective |

Subsequently, all processed B-scans are fed into a well-trained MLFF-Net for defect detection and the predicted health conditions are shown in Table VI. Except for Tree 4, all predictions of Tree 1, 2, and 3 are correct even though the scans are conducted from different sides of the tree trunk. The consistent prediction of the same tree is attributed to including the scans from different sides of the tree trunk when preparing the training dataset. The total false alarm is 18.75% which is satisfactory but slightly higher than the performance reported on the testing dataset measured in the laboratory.

### C. Discussions of Results

The degraded performance of the well-trained MLFF-Net on measurement data of living trees is expected and attributed to

| Scanning position | 1 | 2 | 3 | 4 |
|---|---|---|---|---|
| Tree 1 | Healthy | Healthy | Healthy | Healthy |
| Tree 2 | Healthy | Healthy | Healthy | Healthy |
| Tree 3 | Healthy | Healthy | Healthy | Healthy |
| Tree 4 | Defective | Healthy | Defective | Defective |

two reasons. First, the total amount of scans conducted on living trees is limited due to the difficulties of finding target trees, especially trees with internal defects. As a result, the evaluated accuracy may not be able to reflect the true performance of the model. Second, the MLFF-Net is trained with measurement data of tree trunk samples. Although the tree trunk samples were scanned shortly after being obtained, the variation of the moisture content during the storage could result in different distributions of dielectric properties compared to the living trees, influencing the penetration of the transmitted signal. Additionally, the diversity of trees in terms of the various shapes and individual characteristics of the internal wood structures is not completely included in the current dataset. All of these differences between the tree trunk samples in the dataset and the living trees could lower the model's accuracy when applying it to defect detection in living trees. To address such an issue, transfer learning could be employed to fine-tune the well-trained with another small set of measurement data of the living trees, in which case the weights of the parameters in the model are adjusted to fit the signatures of defects inside the living trees. It should be noted that the proposed deep learning-augmented radar scheme is a data-driven approach, the performance of which could be enhanced by enlarging the dataset size.

The performed on-field test validates the accuracy and effectiveness of the proposed system for defect detection, which can automatically perform the data collection, processing, and prediction within minutes. With adequate training by datasets of live trees that contain diverse defects in the future, the proposed system is capable of rapid and accurate detection of defects inside trees with differing species.

## VIII. CONCLUSION

This work proposed a deep learning-augmented stand-off radar scheme for real-time detection of the defects inside real tree trunks. Firstly, the proposed scheme performs the measurement on the tree trunks in a contactless manner while its antenna is moved on a straight trajectory (on a motorized slider). Then the measured B-scan is processed by a proposed signal processing framework, consisting of free-space response removal, C3-based zero-gating algorithm, and FIR filtering. The framework automatically removes the clutter due to the antenna's internal reflection, and the reflection from the air-bark interface, and suppresses the high-frequency components due to their weak penetration into the tree trunks. By doing so, the proposed signal processing framework improves the SCNR of measurement data of tree trunk samples and living trees by more than 30 dB and 22 dB, respectively, making the defect features more visible in the B-scan. Finally, the processed B-scan is inputted to MLFF-Net, particularly designed for detecting defects of various sizes, positions, and types inside the fresh-cut tree trunk samples with over 96% accuracy. Besides the multilevel feature extraction and fusion, the attention mechanism is included through CAM, which emphasizes contributions of effective features along both channel and spatial dimensions and significantly enhances the network's performance. The robustness of the proposed network has been validated through the ablation study and the comparative study with existing DL techniques. The on-field tests of the proposed radar scheme demonstrated its accuracy and potential for incorporation into the workflow for routine structural health screening of whole tree populations.

It is worth noting that the proposed scheme is validated on the tree species with regularly convex shapes. Applying such a scheme to more tree species with irregular non-convex shapes requires further developments in both air-bark clutter removal and the design of more robust neural networks. Moreover, the MLFF-Net is trained by the measurement dataset of tree trunk samples, the permittivity and conductivity of which are lower compared to that of the living trees due to the relatively low water content. Such a variation could degrade the performance when applying the well-trained model directly to the measurement data of the living tree. Thus, fine-tuning the model with another dataset of living trees is necessary to ensure a satisfied performance. Last but not least, the detection capability of the proposed scheme might not be valid for tree species with much higher relative permittivity and conductivity compared to the Angsana tree, which is attributed to the degraded penetration of the transmitted signal into the tree trunks. Studies, including the design of an antenna with a higher gain that transmits the signal with higher intensity, might mitigate this limitation. Our ongoing work will further investigate new and novel network architectures for the parameter estimation of defects. At the same time, this work will be supplemented with the concurrent development of techniques for imaging the tree interiors using the proposed stand-off radar scheme.